\documentclass[preprint]{vgtc}

\usepackage{times}

\usepackage{mathptmx} 

\onlineid{0}

\vgtccategory{Research}

\vgtcinsertpkg

\preprinttext{\parbox{0.9\textwidth}{$\copyright$ \the\year~IEEE\@. This is the author's version of the article that has been published in the adjunct proceedings of IEEE ISMAR 2024 conference. The final version of this record is available at: \url{https://ieeexplore.ieee.org/document/10765333}.}}

\usepackage{mwe}
\usepackage{enumitem}
\usepackage{xspace}
\usepackage{multirow} 
\usepackage{caption}
\usepackage{subcaption}
\usepackage{tablefootnote}
\usepackage{threeparttable}
\usepackage{tabularx} 
\usepackage{booktabs} 
\usepackage{CJKutf8}

\newcommand{\company}{FieldFusion Ltd.}
\newcommand{\Description}[1]{}

\title{Be There, Be Together, Be Streamed! AR Scenic Live-Streaming for an Interactive and Collective Experience}

\author{Zeyu Huang\thanks{e-mail: zhuangbi@connect.ust.hk}\\ %
    \parbox{1.6in}{\scriptsize \centering The Hong Kong University of Science and Technology, China} %
\and Zuyu Xu\thanks{e-mail: zxubo@connect.ust.hk}\\ %
\parbox{1.6in}{\scriptsize \centering The Hong Kong University of Science and Technology, China} %
\and Yuanhao Zhang\thanks{e-mail: yzhangiy@connect.ust.hk}\\ %
\parbox{1.6in}{\scriptsize \centering The Hong Kong University of Science and Technology, China} %
\and Chengzhong Liu\thanks{e-mail: chengzhong.liu@connect.ust.hk}\\ %
\parbox{1.6in}{\scriptsize \centering The Hong Kong University of Science and Technology, China} %
\and Yanwei Zhao\thanks{e-mail: 458815330@qq.com}\\ %
     \parbox{1.6in}{\scriptsize \centering Zhejiang University, China} %
\and Chuhan Shi\thanks{e-mail: chuhanshi@seu.edu.cn}\\ %
     \scriptsize Southeast University, China %
\and Jason Chen Zhao\thanks{e-mail: zhaoch21@gmail.com}\\ %
     \scriptsize FieldFusion Ltd., China %
\and Xiaojuan Ma\thanks{e-mail: mxj@cse.ust.hk}\\ %
\parbox{1.6in}{\scriptsize \centering The Hong Kong University of Science and Technology, China}}

\teaser{
  \centering
  \includegraphics[width=0.8\linewidth]{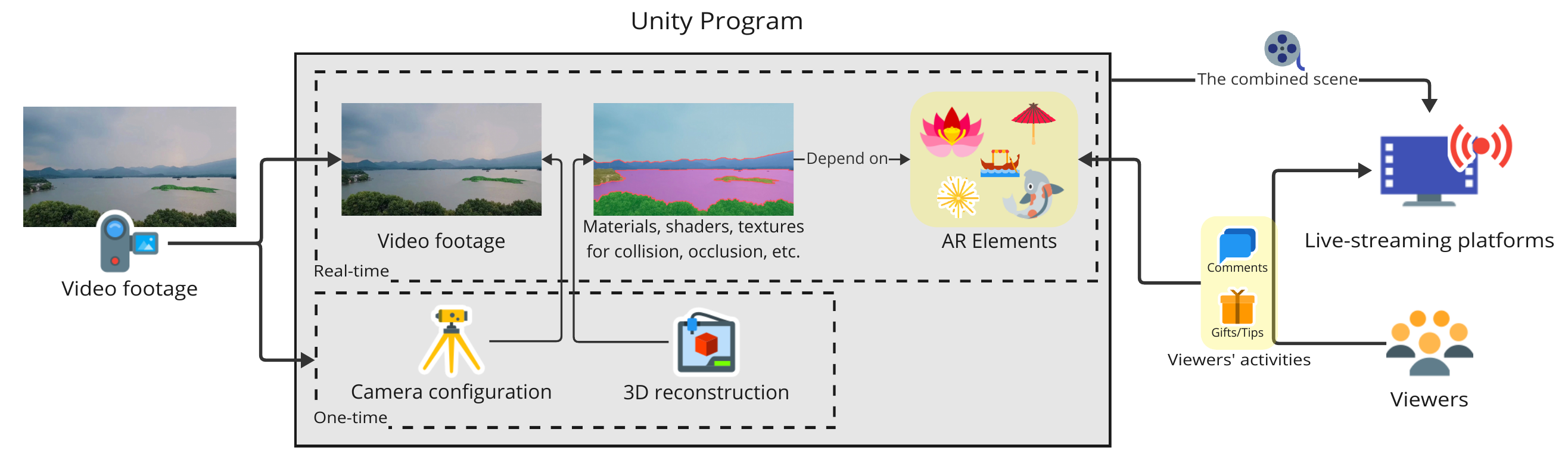}
  \caption{The ARSLS system diagram.}
  \label{fig:diagram}
}
\abstract{%
Scenic Live-Streaming (SLS), capturing real-world scenic sites from fixed cameras without streamers,
combines scene immersion and the social and real-time characteristics of live-streaming into a unique experience.
However, existing SLS affords limited audience interactions to engage them in a collective experience compared to many other live-streaming genres.
It is also difficult for SLS to recreate important but intangible constituents of in-person trip experiences, such as cultural activities.
To offer a more interactive, engaging, and meaningful experience, we propose ARSLS (Augmented Reality Scenic Live-Streaming).
Culturally grounded AR objects with awareness of the live-streamed environment can be overlaid over camera views
to provide additional interactive features while maintaining consistency with the live-streamed scene.
To explore the design space of this new medium, we developed an ARSLS prototype for a famous landscape in China.
A preliminary study (N=15) provided initial insights for ARSLS design.
}


\begin{document}
\maketitle

\section{Introduction}

Live-streaming is a popular means of entertainment nowadays
where people gather together to watch the same video stream at the same time.
Such a time-synced and collective nature often brings about
something beyond pure video-watching experience ---
a collective experience created by spontaneous interactions among viewers in the same live-streaming room~\cite{luYouWatchYou2018,hamiltonStreamingTwitchFostering2014}.
In recent years, a new form of live-streaming called \emph{scenic live-streaming} (SLS)\footnote{Other names of this medium include but are not limited to \emph{world cameras} and \emph{live webcams}.} has emerged:
using fixed cameras to capture real-world environments, often scenic views, without human streamers to create a 24/7, scenery-oriented experience.
SLS appears on both dedicated websites like SkylineWebcams\footnote{\url{https://www.skylinewebcams.com/}}
and general-purpose video platforms like YouTube\footnote{An example of live-streamed Venice: \url{https://www.youtube.com/watch?v=HpZAez2oYsA}}.
The situated experience in a remote location provides an alternative to in-person trips~\cite{jarrattExplorationWebcamtravelConnecting2021},
including having a taste of new places~\cite{jarrattExplorationWebcamtravelConnecting2021, koskelaOtherSideSurveillance2006},
relieving nostalgia for familiar places~\cite{gonzalezKeepingStrongConnections2007},
and gain educational insights~\cite{jarrattExplorationWebcamtravelConnecting2021}.

Apart from the benefits mentioned above,
SLS also allows viewers to participate in real-time social interaction.
Social connection and community building are common purposes people choose live streams over non-interactive videos~\cite{wulfWatchingPlayersExploration2020},
and such goals are realized through the exchange of thoughts and feelings in live chats and other channels on the platform~\cite{huWhyAudiencesChoose2017}.
SLS naturally inherits this trait compared to pre-recorded scenic videos.
Nevertheless, social interactions and collective experiences in SLS are less engaging because there are no human streamers who constantly ``stir up the atmosphere'' for more interactions among the audience~\cite{hamiltonStreamingTwitchFostering2014} and
timely respond to the audience's interactions~\cite{luYouWatchYou2018, haimsonWhatMakesLive2017, liSystematicReviewLiterature2020, zhouMagicDanmakuSocial2019,kimWhoWillSubscribe2019}.
Contrary to the current limitation, we anticipate an even larger potential of SLS in creating unique collective experiences that transcend sole text communications among viewers:
a scenery often bears profound cultural values,
and many of them are reflected through collective cultural activities.
Recreating such activities will not only help people gain a comprehensive understanding of the live-streamed site, but also leverage the scenery itself and the underlying culture to offer unique real-time experiences.

To this end, we propose a medium called \emph{Augmented Reality Scenic Live-Streaming} (ARSLS),
where additional virtual content is
superimposed and integrated into the real-world scene
to actualize proper audience interactions in scenic-oriented live streams.
AR has already been a valid solution to enhance in-person, single-user experiences with real-world scenery~\cite{wangIntangibleCulturalHeritage2018, tungAugmentedRealityMobile2015},
and previous work has also shown the technical feasibility of
applying AR to fixed cameras~\cite{szentandrasiPOSTERINCASTInteractive2015,liuAugmentedRealityassistedIntelligent2017}.
For ARSLS, the AR content shall be aware of the scene's structures and boundaries,
be able to interact with the real-world scene~\cite{azumaSurveyAugmentedReality1997},
and respond to viewers' interactions in the live-streaming room, such as comments and tipping, in real-time.

It is worth noting that the additional AR content and interactivity may largely differ from current SLS --- where the streamed view is mainly for passive appreciation and separate from viewers' interactions.
Hence, we propose ARSLS as a new medium derived from SLS, not a replacement for existing SLS\@.
Focusing on the limited engagement and the cultural activities as two underexploited elements,
the aim of ARSLS is to offer a real-time immersive experience combining scenic, social, and cultural emphases.

To explore the design space of ARSLS, we developed a prototype in a real-world setting and obtained preliminary insights from a qualitative study (N=15).
The ARSLS prototype is set up at West Lake, a famous scenic spot in Hangzhou, China
with intriguing aesthetic and profound cultural values~\cite{ zhangCulturalLandscapeMeanings2019}.
We designed several concrete interactive features,
where viewers' spontaneous and collective interactions are driven by the scenic and cultural elements of the site.
The implementation is in partnership with a local corporation, \company{}
\section{Related Work}

\subsection{AR on Static Cameras}
Apart from the mainstream AR application on mobile devices and Head-Mounted Displays,
AR on static camera images or video footage is another approach
to delivering virtual content in a given scene.
This is made possible by remote rendering techniques~\cite{lambertiStreamingBasedSolutionRemote2007,luRenderFusionBalancingLocal2023}.
Szentandrási et al.~\cite{szentandrasiPOSTERINCASTInteractive2015} applied real-time computed, context-aware AR to video footage captured by surveillance cameras,
and Liu et al.~\cite{liuAugmentedRealityassistedIntelligent2017}
integrated interactive AR into physical machine tools and machining processes.
However, current works mainly focus on technological feasibility and less on designing an engaging and immersive experience.
Our design specifically highlights AR for an engaging SLS experience, including strengthening the connection between the audience and the place, as well as viewers' collective interactions.

\subsection{Interactive Live-Streaming}
Live-streaming is a new form of multimedia service that has gained increasing popularity in recent years.
On modern live-streaming platforms, users can not only watch the stream passively, but also interact with the streamer, other viewers, or the streamed content through live chat (i.e.\ commenting) and gifting (i.e.\ tipping or rewarding)~\cite{luYouWatchYou2018, haimsonWhatMakesLive2017, liSystematicReviewLiterature2020, zhouMagicDanmakuSocial2019, kimWhoWillSubscribe2019}.
Such interactions offer emotional support,
make viewers noticeable to the public,
and help them develop social connections~\cite{luYouWatchYou2018, liSystematicReviewLiterature2020}.

However, most research examines the interactions of more popular streaming categories, e.g., gaming and life sharing~\cite{luYouWatchYou2018}.
SLS, on the other hand, mainly displays real-time views with limited or no motivation for interactions~\cite{jarrattExplorationWebcamtravelConnecting2021, koskelaOtherSideSurveillance2006}.
Thus, the ultimate goal of ARSLS is to provide incentives and opportunities for a comparably interactive viewer experience with other live-streaming categories.

\section{ARSLS Design}

\subsection{System Overview}

As illustrated in \cref{fig:diagram},
the ARSLS is implemented as a Unity program.
The video footage is streamed to the program,
and the perspective of the Unity scene is adjusted
according to the camera device that captures the video.
Since the camera is fixed (hence so is the view),
we can manually reconstruct the 3D structure of the scene in Unity.
Proper shaders and textures are attached to the nearer geometries
of the 3D reconstruction
so that they appear transparent but can occlude virtual objects behind them.
Hitboxes are also configured, as well as the special ripple effects of the water surface whenever virtual objects are in touch with it.
The Unity program also subscribes to the commenting and gifting events in a specific live-streaming room via public API\@.
As such, additional AR objects not only conform to the scene's geometric structure,
but also programmatically respond to viewers' actions in real-time.
Finally, the combined scene of the Unity program is remote-rendered on a server and streamed to the live-streaming room.

Our implementation of the ARSLS prototype was in collaboration with a local company, \company{}
The company implemented the fundamental Unity program that captured user action events, created necessary shaders and art assets, and provided the deployment server.
On top, we built the 3D reconstruction of the West Lake scene,
and designed and implemented the following features that leveraged the real-time commenting and gifting data.

\subsection{The Live-Streamed View}\label{sec:design:site-user}

\begin{figure*}[ht]
  \centering
  \hfill
  \begin{subfigure}[t]{0.46\textwidth}
      \centering
      \includegraphics[width=\textwidth]{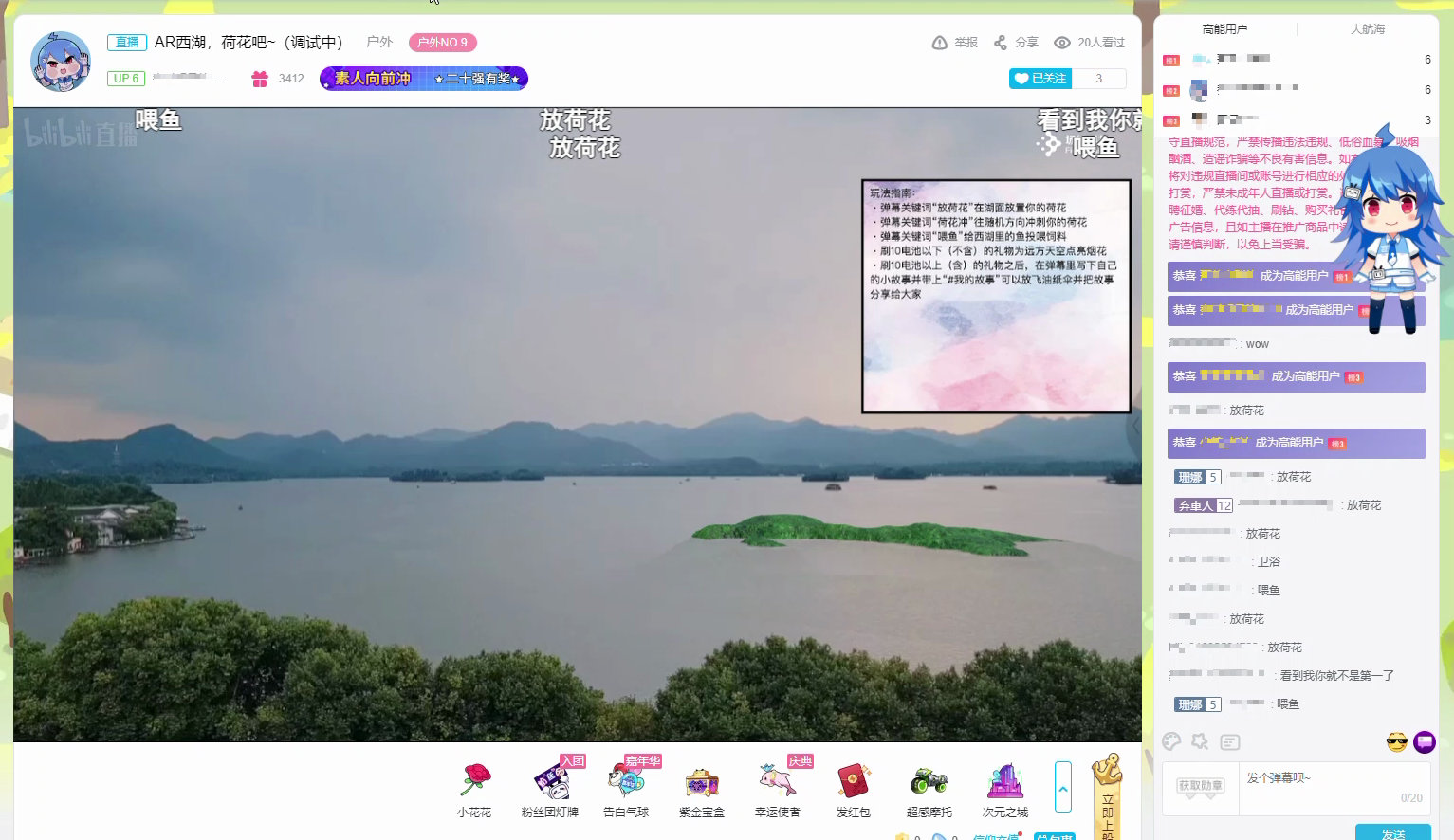}
      \caption{The streaming room and the plain scene before any AR object was created.}
      \label{fig:design:scene}
  \end{subfigure}
  \hfill
  \begin{subfigure}[t]{0.48\textwidth}
      \centering
      \includegraphics[width=\textwidth]{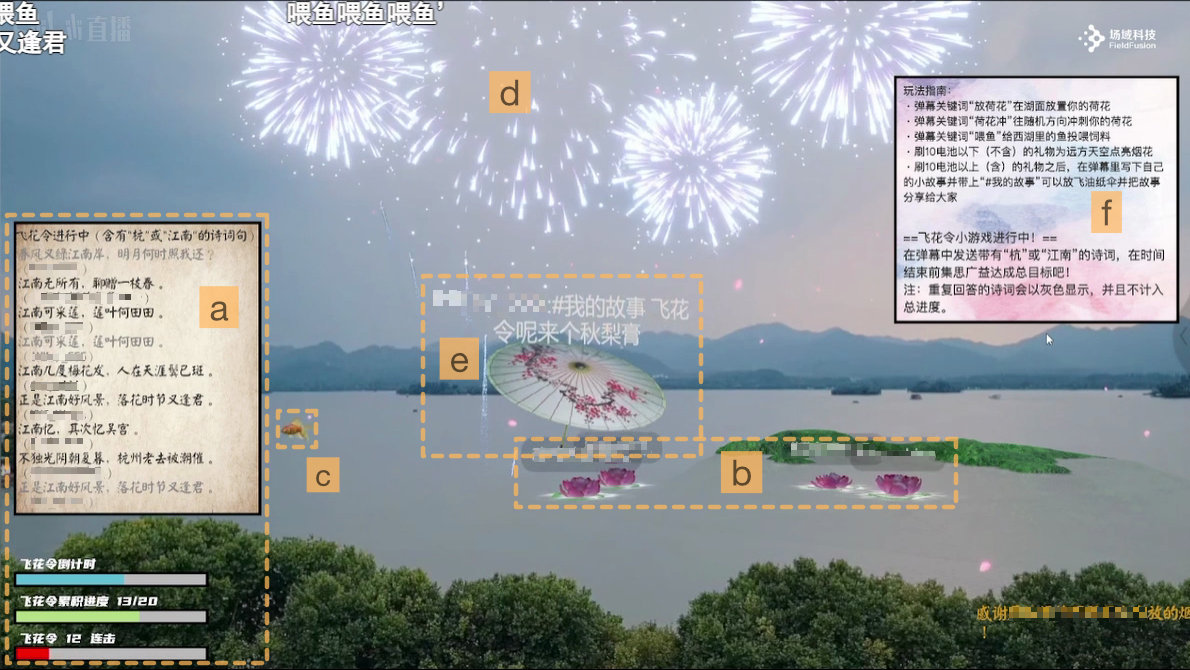}
      \caption{A screenshot at the middle of the session when all features are triggered. \textcircled{a} Chinese verse game ``Fei Hua Ling''; \textcircled{b} Lotus spreading; \textcircled{c} Fish feeding; \textcircled{d} Fireworks; \textcircled{e} Oiled-paper umbrella with stories; \textcircled{f} Instructions.}
      \label{fig:design:features}
  \end{subfigure}
  \caption{The ARSLS design. The features aim to recreate in-person collective activities related to local cultures. Usernames are occluded.}
  \label{fig:design}
\end{figure*}

The West Lake in Hangzhou, China, is one of the best-known scenic sites nationwide.
It is also listed as UNESCO's World Heritage~\cite{StateConservationProperties2019}.
It bears not only the aesthetic value of the landscape with bridges, causeways, pavilions, temples, etc.\@ but also profound historical and cultural values as the theme of countless traditional artworks and poetry~\cite{zhangCulturalLandscapeMeanings2019},
making it a suitable ARSLS site.
Existing city SLS services,
e.g., SkylineWebcams\footnotemark[2] and Live China\footnote{\url{https://livechina.cctv.com}},
often adopted a top-down camera view that overlooks the landscape.
Therefore, we set up a camera that could capture West Lake and the hills far away and on the left from a lookout point (see \cref{fig:design:scene}).

\subsection{AR Features}\label{sec:features}

\subsubsection{Chinese Verse Game ``Fei Hua Ling''}\label{sec:features:feihua}

Given the rich collection of classical Chinese poems about West Lake,
we adopted a traditional competitive literary game about verses in poems called ``Fei Hua Ling'' (\begin{CJK*}{UTF8}{gbsn}飞花令\end{CJK*} in Chinese, lit.\@ ``flying flower command'') with slight modifications to suit the live-streaming context.
At the beginning of the game, a keyword (e.g., ``flower'', ``snow'', and ``wine'') or a theme (e.g., nostalgia) is announced.
The audience collaboratively quotes verses in live comments that contain the keyword.
They win the game together if the total number of unique, valid verses exceeds a threshold within a given time (i.e., 5 minutes in our implementation).
A virtual game board on the left side shows
the verse keyword,
the most recent nine valid verses mentioned by viewers,
the countdown, the combo, and the total progress.
With Fei Hua Ling, we aimed to
(1) engage the audience in collective efforts~\cite{haimsonWhatMakesLive2017, webbDistributedLivenessUnderstanding2016}
and
(2) encourage them to recall and build connections with the local culture depicted in classical literature.
We designed two rounds of Fei Hua Ling, 10 minutes apart.
The first round uses the most conventional keyword, ``flower'' (the origin of the game's name).
If the audience wins,
they will unlock a lasting effect of flower petals flying all over the sky.
The second round prompts a theme of ``Hangzhou (the city where West Lake is) and Jiangnan (lit. `South of the Yangtze River', a geographic area where Hangzhou resides)''.
If the audience wins, a bunch of colorful fireworks will set off,
creating a beautiful view.

\subsubsection{Lotus Spreading} \label{feat:lotus-spreading}

Releasing water lanterns is another traditional collective activity\footnote{\url{https://en.wikipedia.org/wiki/Water_lantern}}.
In Chinese culture, a water lantern is often in the form of a lotus flower.
Thus, we mimic this collective cultural activity in ARSLS\@.
Viewers can place their own virtual lotuses on the lake by issuing a specific comment, \texttt{release my lotus}.
The lotus displays the user's ID above it, slowly drifting to the right to prevent the scene from being completely occluded.
Each user can only have one lotus on the scene;
however, if the lotus goes beyond the screen,
it will be destroyed, and the user can spawn another one.
While the lotus drifts, it triggers virtual ripple effects on the lake.
Users can send \texttt{dash my lotus} to make their lotus dash in a random direction.

Apart from the ornamental value of lotuses and the cultural meaning of this activity,
the feature was also aimed to reinforce viewers' sense of participation
through co-creating a shared view~\cite{gonzalezKeepingStrongConnections2007, dengBlendedTourismExperiencescape2019, hamiltonStreamingTwitchFostering2014}.
The unique object association might turn
lotuses into virtual embodiments of the audience members~\cite{genayBeingAvatarReal2022},
hence offering a proxy for interpersonal interactions
to meet audiences' social needs in live-stream sessions~\cite{webbDistributedLivenessUnderstanding2016}.

\subsubsection{Fish Feeding}

In the live-streaming, we designed fish feeding as a basic interaction with minimal cognitive effort.
When users sent the comment \texttt{feed fish}, fish food was dropped at a random location over the lake.
Then, a virtual fish jumped out of the water to consume the food with a water-splashing sound.
The fish picked up a random look from common koi and goldfish in China.
This feature was for viewers to familiarize themselves with the ARSLS medium.

\subsubsection{Fireworks}

Fireworks were ancient Chinese inventions
and often launched in batches, with spectators gathering together as a collective activity.
In addition, people who launch fireworks first are considered to have more luck~\cite{timothys.y.lammuseumofanthropologyChineseNewYear2021},
which adds nuanced social indication to this activity.
Therefore, we associated a fireworks feature with viewers' tipping action.
Participants often expect acknowledgment when they tip in a live stream,
as it can provide a sense of satisfaction and privilege~\cite{zhouMagicDanmakuSocial2019, luYouWatchYou2018}.
Apart from verbal thanks, users may also expect that their ideas are conveyed exclusively, or that other audiences appreciate the outcome of their tipping~\cite{luYouWatchYou2018, kimWhoWillSubscribe2019}.
As a one-time response to users' cheap gifts (below 10 CNY),
virtual fireworks are set off from the far end of the scene and explode in the sky.
The tipper's name will be displayed in the meantime.
This way, the scene recognizes the monetary input with delicate decorations and a similar sense of privilege.

\subsubsection{Oiled-Paper Umbrella with Stories}

Previous studies~\cite{bennettTellingTalesNostalgia2009, devineRemovingRoughEdges2014} discovered the emotional value of sharing personal stories,
especially to relieve people's nostalgia and build strong communities.
Considering the symbolization of romance and reunion in Chinese culture of the oiled-paper umbrellas, we combined this imagery with storytelling to help users convey their emotions through AR\@.
Users who send gifts of 10 CNY or above are assigned an oiled-paper umbrella with a random
texture.
They can then type their stories in a live comment with the hashtag \texttt{\#MyStory}.
The umbrella, with the name and words attached, will be generated at the bottom of the scene and slowly fly up to the sky.
We aimed to give tipped users a more eye-catching way to express themselves.
The prompt for personal stories also aimed to create emotional appeals to other audiences.

\section{Preliminary Findings}\label{sec:iter}

A qualitative study (N=15, 9 females and 6 males) was conducted against our ARSLS prototype.
Participants were invited to the streaming room simultaneously and remotely.
They freely experienced ARSLS together for 20 minutes.
All instructions were displayed on the screen;
hence, no prior instructions were given to best simulate real-world live-streaming scenarios.
The two rounds of verse games (see \cref{sec:features:feihua}) were inserted at the third and the eleventh minutes of the session.
We asked participants to interact with the system and use the gifting feature similarly to their regular live-streaming experiences.
Only after the study did we offer to reimburse the money they spent on gifts.
Apart from the reimbursement, each received a stipend of 80 CNY\@.
Once the streaming ended, participants were invited to semi-structured interviews separately to reflect on their general perception of the ARSLS, the interactive feature design, and the integration of AR in the real scene.

One aspect that emerged in the interview was visual diversity and personalization. The lotuses served as virtual proxies of viewers, participants desired their own proxy to ``stand out'' and become ``identifiable on the screen''. Otherwise, visually repetitive objects would cluster when more viewers joined in. Nevertheless, it suggested the affordance of virtual embodiment in our prototype, which can be an effective way of creating stronger connections between viewers and the environment in ARSLS\@.

Some participants have complained about the lack of competitive elements to engage them.
For example, the verse game did not provide enough mechanisms to encourage viewers' participation,
and the lotus movements are too gentle to create an entertaining taste.
The rationale behind can be twofold: some may see competitiveness and entertainment as incentives to join an \emph{interactive} experience, while some --- driven by the perceived virtual embodiment --- desired teasing and rapports with other viewers as a \emph{collective} experience.

Also relevant to previous aspects was the emphasis of identifiability of their input, and this demand was intensified by the innate delay of live streams. Especially for the fish feature, when people issued a command and did not know which fish is triggered by them, their interest in using the same feature again would diminish.

The last aspect is vicariousness and immersion.
Unlike regular SLS, the strengthened sense of virtual embodiment may have imposed stricter design considerations for this aspect. In our case, viewers were implicitly guided to immerse in first-person views, so the overlooking scenic view fell short.
They could not identify representative landmarks in recognizable sizes, nor could they relate themselves to such a view. Consequently, they could not feel immersed in or connected to the scene.

Overall, the relationship between virtual embodiment and immersion presents a major design challenge. Both elements are crucial for a successful ARSLS experience, yet they must be carefully balanced. The virtual embodiment recovers social interactions --- an integral part of live streams --- in a playful way with no facilitators. The immersion ensured enough significance of the scene, so that viewers do not simply join a game, but can obtain personal, educational, or other values from the scene.

\section{Conclusion and Future Work}

Through our prototype and the preliminary results,
we have identified participants' intertwined desire for an engaging, embodying collective activity and towards a vicarious immersion.
This aligns with, hence generally verifies, our proposal of ARSLS as a combination of scenic and interactive experiences.
However, it also indicated a complex design process to balance every aspect.
In the future,
we would like to respond to such tensions in ARSLS designs via continuous refinement and evaluation of the prototype.


\bibliographystyle{abbrv-doi}
\bibliography{main}






\end{document}